\documentclass[12pt,preprint]{aastex}

\usepackage{epsfig}
\usepackage{graphics}

\newcommand{\tn}[1]{\textnormal{#1}}
\newcommand{\fltr}[2]{$\tn{#1}_{\tn{\tiny{#2}}}$}

\shorttitle{GRB Afterglows data set}
\shortauthors{Kann, Zeh, Klose}

\begin{document}

\title{A catalog of optical/near-infrared data on GRB afterglows in the
       pre-\emph{Swift} era.\\ I. Light curve information}

\author{D. A. Kann,\altaffilmark{1}
A. Zeh,\altaffilmark{1}
S. Klose\altaffilmark{1}}

\altaffiltext{1}{Th\"uringer Landessternwarte Tautenburg, Sternwarte 5,
D--07778 Tautenburg, Germany}

\begin{abstract}

The present catalog is the result of our attempts to collect all published
photometric data on GRB afterglows observed in the pre-\emph{Swift} era by the
end of 2004 in order to gain statistical insight on the phenomenology of GRB
afterglows. Part I contains all \emph{published} data on GRB afterglows in filters
we used in Zeh, Klose, \& Kann (2005) to create reference light curves and
derive light curve parameters (mostly $R$ band, but a few bursts have better
data in other colors) including the corresponding references.  The catalog
includes GCN data as well as data published in refereed journals. No data have
been omitted or evaluated in any way (with the exception of a very small
number of data that turned  out to be not related to an afterglow).  Part II
will contain color information via the observed light curves in the other
photometric bands (Kann et al. 2006, in preparation).  Using a simple computer
program that can handle strings  our catalog is easy to use since all
tables are provided in {\TeX} format. For an on-line searchable GCN catalog, we
refer to the work done by Quimby et al. (2003).

Our catalog includes photometric data on 59 bursts (GRB 970228 - GRB 041006)
with altogether 4883 data points. Most data are from GRB 030329 (2759 data
points), followed by GRB 021004 (393 data points), while 13 bursts have less
than 10 data points. In the case of GRB 030329 we have not included the
extensive data list which is on-line provided by Lipkin et al. (2004).

The catalog is organized as follows. We provide for every GRB the reported
time of the photometric measurement after the burst in units of days,  the
corresponding magnitude, the 1$\sigma$ error, the date of the measurement, and
information on whether the reported data are corrected for Galactic extinction (NE = not
corrected, CE = corrected) or host contribution (NH = not corrected, CH = corrected).
References are given for every individual burst. Remarks
are indicated by numbers in the last column. The time after the burst was
calculated based on the information given in the corresponding \emph{BACODINE}
messages. We acknowledge here the work done by Barthelmy et al. (2001) in
collaboration with all teams that operated the various satellites/GRB
experiments.

We intend to update this catalog as soon as more data on pre-\emph{Swift}
bursts become public. If any researchers have photometric data that they
do not plan to publish in a refereed journal but wish to share with the community,
we will gladly include it in this database. Please contact kann@tls-tautenburg.de.

When making use of the present catalog for a publication, please make a
reference to Zeh et al. (2005).

\vspace{1cm}
\end{abstract}


\hspace*{-10mm} \bf Acknowledgments:\rm \small \\
A.Z. and S.K. acknowledge financial support by DFG grant Kl
766/11-1. D.A.K. and S.K. acknowledge financial support by DFG grant Kl
766/13-2. We wish to thank S. Barthelmy, NASA, for the upkeep of the GCN
Circulars and the \emph{BACODINE} messages
and J. Greiner, MPE Garching, for the great "GRB Big Table" (Greiner 2005).


\newpage

\begin{deluxetable}{lc|lc|lc|lc}
\tablecaption{\bf{Number of data points per GRB}}
\renewcommand{\tabcolsep}{8pt}
\tablehead{
\colhead{GRB}  &
\colhead{\#} &
\colhead{GRB}  &
\colhead{\#} &
\colhead{GRB}  &
\colhead{\#} &
\colhead{GRB}  &
\colhead{\#}}
\startdata
GRB 970228 & 46 &
GRB 970508 & 91 &
GRB 970815 & 7 &
GRB 971214 & 15 \\
GRB 980326 & 26 &
GRB 980329 & 36 &
GRB 980519 & 35 &
GRB 980613 & 13 \\
GRB 980703 & 27 &
GRB 990123 & 118 &
GRB 990308 & 6 &
GRB 990510 & 66 \\
GRB 990705 & 3 &
GRB 990712 & 25 &
GRB 991208 & 24 &
GRB 991216 & 54 \\
GRB 000131 & 4 &
GRB 000301C & 62 &
GRB 000418 & 28 &
GRB 000630 & 11 \\
GRB 000911 & 27 &
GRB 000926 & 61 &
GRB 001007 & 10 &
GRB 001011 & 6 \\
GRB 010222 & 89 &
GRB 010921 & 30 &
GRB 011121 & 30 &
GRB 011211 & 49 \\
GRB 020124 & 21 &
GRB 020305 & 14 &
GRB 020322 & 10 &
GRB 020331 & 16 \\
GRB 020405 & 40 &
GRB 020410 & 4 &
GRB 020813 & 67 &
XRF 020903 & 17 \\
GRB 021004 & 393 &
GRB 021211 & 73 &
GRB 030115 & 7 &
GRB 030131 & 7 \\
GRB 030226 & 51 &
GRB 030227 & 19 &
GRB 030323 & 47 &
GRB 030324 & 7 \\
GRB 030328 & 25 &
GRB 030329 & 2759 &
GRB 030418 & 28 &
GRB 030429 & 19 \\
GRB 030528 & 5 &
XRF 030723 & 45 &
GRB 030725 & 17 &
GRB 031203 & 32 \\
GRB 031220 & 8 &
GRB 040106 & 7 &
GRB 040422 & 3 &
GRB 040827 & 13 \\
XRF 040916 & 14 &
GRB 040924 & 23 &
GRB 041006 & 93 &
&\\
\enddata
\end{deluxetable}

\newpage


\begin{deluxetable}{lllllllll}
\tablecaption{\bf{GRB 970228}}
\tablehead{
\colhead{$dt$}  &
\colhead{mag} &
\colhead{+error} &
\colhead{-error} &
\colhead{band} &
\colhead{date} &
\colhead{corrections} &
\colhead{reference} &
\colhead{remark}}
\startdata
0.64    &   21.1    &   0.2 &   0.2 &   \fltr{R}{C} &   970228.77   &   NE,NH   &   Gal387  &   \\
0.68    &   20.5    &   0.5 &   0.5 &   \fltr{R}{C} &   970228.81   &   NE,NH   &   Gal387  &   \\
0.686   &   20  &   0.5 &   0.5 &   \fltr{R}{C} &   970228.81   &   NE,NH   &   Gal536  &   \\
0.703   &   21.1    &   0.2 &   0.2 &   \fltr{R}{C} &   970228.827  &   NE,NH   &   Gua328  &   \\
0.706   &   20.58   &   0.28    &   0.28    &   \fltr{R}{C} &   970228.83   &   NE,NH   &   Gal536  &   \\
0.87    &   20.9    &   0.14    &   0.14    &   \fltr{R}{C} &   970228.99   &   NE,NH   &   Gal387  &   4\\
1.667   &   $>$21.4 &   0   &   0   &   \fltr{R}{C} &   970301.791  &   NE,NH   &   Gua328  &   \\
2.98    &   22.2    &   0.7 &   0.7 &   \fltr{R}{C} &   970303.10   &   NE,NH   &   Gal387  &   4\\
\enddata
\tablerefs{$\ $\\ $\ $\\
 Cas523 = Castander, F. J. et al. 1999, ApJ, 523, 593\\
 Gal387 = Galama, T. J. et al. 1997, Nature, 387, 479\\
 Gal536 = Galama, T. J. et al. 2000, ApJ, 536, 185\\
 Gua328 = Guarnieri, A. et al. 1997, A\&A, 328, L13\\
 van386 = van Paradijs, J. et al. 1997, Nature, 386, 686\\
 Wij288 = Wijers, R. A. M. J. et al. 1997, MNRAS, 288, L51\\
 }
\tablecomments{$\ $\\ $\ $\\
This afterglow had a very close neighbor, an M dwarf $2.\!\arcsec9$ away.
Light from this star (\fltr{R}{C} = 21.906$\pm$0.067 (NE; Gal536)) may
contaminate the afterglow magnitudes (e.g. Gua328).\\
1 = host galaxy with optical transient flux removed\\
2 = from stacked HST images, taken 970326 and 970407\\
3 = time replaced with time from van386\\
4 = converted into \fltr{R}{C}\\
5 = converted from published flux density in $\mu$Jy and re-reddened\\
}
\end{deluxetable}
\begin{deluxetable}{llllllll}
\tablecaption{\bf{GRB 970508}}
\tablehead{
\colhead{$dt$}  &
\colhead{mag} &
\colhead{error} &
\colhead{band} &
\colhead{date} &
\colhead{corrections} &
\colhead{reference} &
\colhead{remark}}
\startdata
0.224       & 21.2  & 0.1   & \fltr{R}{C}   &  970509.128    & NE,NH & Cas279   &     \\
0.291       & 21.08 & 0.15  & \fltr{R}{C}   &  970509.195    & NE,NH & Gar500   & 3    \\
0.296       & 21.25 & 0.05  & \fltr{R}{C}   &  970509.20      & NE,NH & Gal497   &     \\
0.841       & 21.19 & 0.25  & \fltr{R}{C}   &  970509.75     & NE,NH & Sok334   &     \\
0.944       & 21.13 & 0.18  & \fltr{R}{C}   &  970509.85     & NE,NH & Sok334   &     \\
0.995       & 20.9  & 0.1   & \fltr{R}{C}   &  970509.899    & NE,NH & Cas279   &     \\
0.995       & 20.7  & 0.1   & \fltr{R}{C}   &  970509.899    & NE,NH & Gar500   &     \\
1.026       & 20.88 & 0.05  & \fltr{R}{C}   &  970509.93     & NE,NH & Gal497   &     \\
\enddata
\tablerefs{$\ $\\ $\ $\\
 Blo507 = Bloom, J. S. et al. 1998, ApJ, 507, L25\\
 Cas279 = Castro-Tirado, A. J. et al. 1998, Science, 279, 1011\\
 Djo387 = Djorgovski, S. G. et al. 1997, Nature, 387, 876\\
 Gal497 = Galama, T. J. et al. 1998, ApJ, 497, L13\\
 Gar500 = Garcia, M. R. et al. 1998, ApJ, 500, L105\\
 Kel4496 = Kelemen, J. et al. 1997, IBVS 4496\\
 Ped496 = Pedersen, H. et al. 1998, ApJ, 496, 311\\
 Pia492 = Pian, E. et al. 1998, ApJ, 492, L103\\
 Sok334 = Sokolov, V. V. et al. 1998, A\&A,  334, 117\\
 Sok344 = Sokolov, V. V. et al. 1999, A\&A,  344, 43\\
 Zha337 = Zharikov, S. V. et al. 1998, A\&A,  337, 356\\
 }
\tablecomments{$\ $\\ $\ $\\
The data point from IBVS 4594 (M. Nalezyty, 1998) could not be included
as no UT time is given.\\
1 = re-reddened according to the values given in Djo387\\
2 = derived from HST STIS Clear (V, \fltr{I}{C}) and HST NICMOS F160W (H)\\
3 = converted from Djo387 Gunn r measurements\\
4 = dt and date according to Gar500\\
5 = also presented in Zha337\\
}
\end{deluxetable}
\begin{deluxetable}{llllllll}
\tablecaption{\bf{GRB 970815}}
\tablehead{
\colhead{$dt$}  &
\colhead{mag} &
\colhead{error} &
\colhead{band} &
\colhead{date} &
\colhead{corrections} &
\colhead{reference} &
\colhead{remark}}
\startdata
1.695   &   $>$19.9 &       &   \fltr{I}{C} &   970817.2    &   NE,NH   &   Ada6725 &   \\
2.645   &   22.1    &   0.1 &   \fltr{I}{C} &   970818.15   &   NE,NH   &   Sod2837 &   \\
4.775   &   22.3    &   0.1 &   \fltr{I}{C} &   970820.28   &   NE,NH   &   Sod2837 &   \\
5.695   &   $>$20.5 &       &   \fltr{I}{C} &   970821.2    &   NE,NH   &   Ada6725 &   \\
9.695   &   $>$21   &       &   \fltr{I}{C} &   970825.2    &   NE,NH   &   Ada6725 &   \\
11.695  &   $>$21   &       &   \fltr{I}{C} &   970827.2    &   NE,NH   &   Ada6725 &   \\
1837.5  &   $>$24   &       &   \fltr{I}{C} &   020827  &   NE,NH   &   Sod2837 &   \\
\enddata
\tablerefs{$\ $\\ $\ $\\
 Ada6725 = Adams, M. T. et al. 1997, IAUC 6723\\
 Sod2837 = Soderberg, A. M. et al. 2004, GCN 2837\\
 }
\tablecomments{$\ $\\ $\ $\\
The afterglow of this GRB was not found until 2004, when a Chandra observation
(and non-detection) of an X-ray source outside the RXTE/ASM error box lead to the
conclusion that this source was the X-ray afterglow of GRB 970815 (Mirabal, N. et al. 2005, ApJ, 620, 379). The re-examination
of images found the optical afterglow at the same position (Sod2837 supersedes the upper
limits reported in IAUC 6723 by S. R. Kulkarni et al.). Thus, some of the upper limits
are of regions (i. e. the error box) not including the afterglow.
}
\end{deluxetable}
\begin{deluxetable}{lllllllll}
\tablecaption{\bf{GRB 971214}}
\tablehead{
\colhead{$dt$}  &
\colhead{mag} &
\colhead{+error} &
\colhead{-error} &
\colhead{band} &
\colhead{date} &
\colhead{corrections} &
\colhead{reference} &
\colhead{remark}}
\startdata
0.5273         & 21.7    & 0.10    & 0.10 & \fltr{R}{C}      & 971215.50       & NE,NH & Kul393     &   \\
0.5373         & 22.1   & 0.1    & 0.10 & \fltr{R}{C}      & 971215.51       & NE,NH & Kul393     & \\
0.5373         & 22.1   & 0.01    & 0.01 & \fltr{R}{C}      & 971215.51       & NE,NH & Hal393     &    \\
0.5373         & 22.06   & 0.06    & 0.06 & \fltr{R}{C}      & 971215.51       & NE,NH & Die503     &   \\
0.827         & $>$21   & 0    & 0 & \fltr{R}{C}      & 971215.8       & NE,NH & Ito6788     &  \\
1.4873         & 23.3    & 0.17    & 0.17 & \fltr{R}{C}      & 971216.46       & NE,NH & Kul393     &   \\
1.5473         & 23.7    & 0.30    & 0.30 & \fltr{R}{C}      & 971216.52       & NE,NH & Kul393     & 1 \\
1.5573         & 23.36   & 0.13    & 0.13 & \fltr{R}{C}      & 971216.53       & NE,NH & Die503     &   \\
\enddata
\tablerefs{$\ $\\ $\ $\\
 Die503 = Diercks, A. et al. 1998, ApJ, 503, L10\\
 Hal393 = Halpern, J. P. et al. 1998, Nature, 393, 41\\
 Ito6788 = Itoh, N. et al. 1997, IAUC 6788\\
 Kul393 = Kulkarni, S. R. et al. 1998, Nature, 393, 35\\
 Sok372 = Sokolov, V. V. et al. 2001, A\&A, 372, 438\\
 }
\tablecomments{$\ $\\ $\ $\\
1 = also in Hal393
}
\end{deluxetable}
\begin{deluxetable}{lllllllll}
\tablecaption{\bf{GRB 980326}}
\tablehead{
\colhead{$dt$}  &
\colhead{mag} &
\colhead{+error} &
\colhead{-error} &
\colhead{band} &
\colhead{date} &
\colhead{corrections} &
\colhead{reference} &
\colhead{remark}}
\startdata
0.422        & 21.19  & 0.1   & 0.1 & \fltr{R}{C}     & 980327.31      & NE,NH & Gro502 &   \\
0.462        & 21.25  & 0.03   & 0.03 & \fltr{R}{C}     & 980327.35      & NE,NH & Blo401   &   \\
0.513        & 21.98  & 0.16   & 0.16 & \fltr{R}{C}     & 980327.401     & NE,NH & Gro502   &       \\
0.549        & 22.18  & 0.16   & 0.16 & \fltr{R}{C}     & 980327.437     & NE,NH & Gro502   &       \\
0.952        & $>$21.85  & 0   & 0 & \fltr{R}{C}     & 980327.84     & NE,NH & Gro502   &       \\
0.964        & $>$22.0  & 0   & 0 & \fltr{R}{C}     & 980327.852     & NE,NH & Gro502   &       \\
0.98        & 22.8  & 0.7   & 0.7 & \fltr{R}{C}     & 980327.868     & NE,NH & Cas138   &   2   \\
1.128        & 23.66  & 0.12   & 0.12 & \fltr{R}{C}     & 980328.016     & NE,NH & Gro502   &       \\
\enddata
\tablerefs{$\ $\\ $\ $\\
 Blo401 = Bloom, J. S. et al. 1999, Nature, 401, 453\\
 Cas138 = Castro-Tirado, A. J. et al. 1999, A\&AS, 138, 449\\
 Gro502 = Groot, P. J. et al. 1998, ApJ, 502, L123\\
 }
\tablecomments{$\ $\\ $\ $\\
1 = deduced from spectrum\\
2 = time estimated from graphic\\
}
\end{deluxetable}
\begin{deluxetable}{lllllllll}
\tablecaption{\bf{GRB 980329}}
\tablehead{
\colhead{$dt$}  &
\colhead{mag} &
\colhead{error} &
\colhead{band} &
\colhead{date} &
\colhead{corrections} &
\colhead{reference} &
\colhead{remark}}
\startdata
0.6741  &   $>$20   &   0   &   0   &   \fltr{R}{C} &   980329.83   &   NE,NH   &   Pal336  &   \\
0.6941  &   $>$22   &   0   &   0   &   \fltr{R}{C} &   980329.85   &   NE,NH   &   Pal336  &   \\
0.6993  &   $>$22.2 &   0   &   0   &   \fltr{R}{C} &   980329.8552 &   NE,NH   &   Gor347  &   \\
0.7441  &   $>$21   &   0   &   0   &   \fltr{R}{C} &   980329.9    &   NE,NH   &   Pal336  &   \\
0.7441  &   $>$22   &   0   &   0   &   \fltr{R}{C} &   980329.9    &   NE,NH   &   Pal336  &   \\
0.7497  &   23.7    &   0.5 &   0.5 &   \fltr{R}{C} &   980329.9056 &   NE,NH   &   Gor347  &   \\
0.7606  &   $>$18.5 &   0   &   0   &   \fltr{R}{C} &   980329.9165 &   NE,NH   &   Ped52   &   \\
0.7641  &   $>$21   &   0   &   0   &   \fltr{R}{C} &   980329.92   &   NE,NH   &   Pal336  &   \\
\enddata
\tablerefs{$\ $\\ $\ $\\
 Gor347 = Gorosabel, J. et al. 1999, A\&A, 347, L31\\
 Jau402 = Jaunsen, A. O. et al. 2003, A\&A, 402, 125\\
 Pal336 = Palazzi, E. et al. 1998, A\&A, 336, L95\\
 Ped52 = Pedersen, H. et al. 1998, GCN 52\\
 Rei517 = Reichart, D. E. et al. 1999, ApJ, 517, 692\\
 Yos577 = Yost, S. A. et al. 2002, ApJ, 577, 155\\
 }
\tablecomments{$\ $\\ $\ $\\
1 = Transformed from HST STIS 50CCD Clear filter\\
2 = also in Rei517\\
}
\end{deluxetable}
\begin{deluxetable}{llllllll}
\tablecaption{\bf{GRB 980519}}
\tablehead{
\colhead{$dt$}  &
\colhead{mag} &
\colhead{error} &
\colhead{band} &
\colhead{date} &
\colhead{corrections} &
\colhead{reference} &
\colhead{remark}}
\startdata
0.5399        & 19.84  & 0.01   & \fltr{R}{C}     & 980520.054     & NE,NH & Jau546     & \\
0.6369        & 20.27  & 0.01   & \fltr{R}{C}     & 980520.151     & NE,NH & Jau546     & \\
0.6484        & 20.28  & 0.10   & \fltr{R}{C}     & 980520.1625    & NE,NH & Vrb528     & \\
0.6489        & 20.39  & 0.12   & \fltr{R}{C}     & 980520.163     & NE,NH & Hal517     & \\
0.7089        & 20.48  & 0.03   & \fltr{R}{C}     & 980520.223     & NE,NH & Jau546     & \\
0.7148        & 20.68  & 0.14   & \fltr{R}{C}     & 980520.2289    & NE,NH & Vrb528     & \\
0.7149        & 20.77  & 0.15   & \fltr{R}{C}     & 980520.229     & NE,NH & Hal517     & \\
0.7729        & 20.87  & 0.13   & \fltr{R}{C}     & 980520.287     & NE,NH & Hal517     & \\
\enddata
\tablerefs{$\ $\\ $\ $\\
 Hal517 = Halpern, J. P. et al. 1999, ApJ, 517, L105\\
 Jau402 = Jaunsen, A. O. et al. 2003, A\&A, 402, 125\\
 Jau546 = Jaunsen, A. O. et al. 2001, ApJ, 546, 127\\
 Vrb528 = Vrba, F. J. et al. 2000, ApJ, 528, 254\\
}
\tablecomments{$\ $\\ $\ $\\
The GRB 980519 light curve presented in Jau546 contains several late upper
limits that are not given in the data table, two in \fltr{I}{C} ($\approx$16 and
$\approx$40 days after the burst) and one in \fltr{R}{C} ($\approx$105 days after
the burst).\\
1 = transformed from HST STIS 50CCD Clear filter\\
2 = part of merging system in which the GRB happened\\
3 = complete merging system, $1.\!\arcsec88$ aperature\\
}
\end{deluxetable}
\begin{deluxetable}{llllllll}
\tablecaption{\bf{GRB 980613}}
\tablehead{
\colhead{$dt$}  &
\colhead{mag} &
\colhead{error} &
\colhead{band} &
\colhead{date} &
\colhead{corrections} &
\colhead{reference} &
\colhead{remark}}
\startdata
0.688   &   22.81       &   0.15    &   R                               &   980613.89       &   NE,NH   &   Hjo576  &   \\
0.688   &   23.19       &   0.20    &   R                               &   980613.89       &   NE,CH   &   Hjo576  &   \\
0.7     &   $>$20.5 &               &   R                               &   980613.9        &   NE,NH   &   Cas102  &   \\
1.038   &   22.96       &   0.09    &   R                               &   980614.24       &   NE,NH   &   Hjo576  &   \\
1.038   &   23.53       &   0.20    &   R                               &   980614.24       &   NE,CH   &   Hjo576  &   \\
3.10    &   23.3        &               &   R                               &   980616.30       &   NE,NH   &   Djo114  &   \\
3.10    &   24.5        &   0.5     &   R                               &   980616.30       &   NE,CH   &   Djo117  &   \\
4.708   &   $>$24.4 &               &   R                               &   980617.91       &   NE,NH   &   Hjo576  &   \\
\enddata
\tablerefs{$\ $\\ $\ $\\
 Cas102 = Castro-Tirado, A. J. et al. 1998, GCN 102\\
 Djo114 = Djorgovski, S. G. et al. 1998, GCN 114\\
 Djo117 = Djorgovski, S. G. et al. 1998, GCN 117\\
 Djo189 = Djorgovski, S. G. et al. 1999, GCN 189\\
 Hjo576 = Hjorth, J. et al. 2002, ApJ, 576, 113\\
 Ode105 = Odewahn, S. et al. 1998, GCN 105\\
 Sok118 = Sokolov, V. V. et al. 1998, GCN 118\\
 Sok372 = Sokolov, V. V. et al. 2001, A\&A, 372, 438\\
}
\end{deluxetable}
\begin{deluxetable}{llllllll}
\tablecaption{\bf{GRB 980703}}
\tablehead{
\colhead{$dt$}  &
\colhead{mag} &
\colhead{error} &
\colhead{band} &
\colhead{date} &
\colhead{corrections} &
\colhead{reference} &
\colhead{remark}}
\startdata
0.9405        & 21.00  & 0.09   & \fltr{R}{C}     & 980704.123     & NE,NH & Cas511   &    \\
0.9835        & 21.04  & 0.06   & \fltr{R}{C}     & 980704.166     & NE,NH & Cas511   &    \\
1.0055        & 21.13  & 0.05   & \fltr{R}{C}     & 980704.188     & NE,NH & Cas511   &    \\
1.2355        & 21.17  & 0.12   & \fltr{R}{C}     & 980704.418     & NE,NH & Cas511   &    \\
1.2505        & 21.35  & 0.11   & \fltr{R}{C}     & 980704.433     & NE,NH & Cas511   &    \\
1.2655        & 21.22  & 0.10   & \fltr{R}{C}     & 980704.448     & NE,NH & Cas511   &    \\
1.2945        & 21.34  & 0.2    & \fltr{R}{C}     & 980704.477     & NE,NH & Cas511   &    \\
1.2975        & 21.28  & 0.18   & \fltr{R}{C}     & 980704.48      & NE,NH & Blo508   &   \\
\enddata
\tablerefs{$\ $\\ $\ $\\
 Bar133 = Bartolini, C. et al. 1998, GCN 133\\
 Blo508 = Bloom, J. S. et al. 1998, ApJ, 508, L21\\
 Blo702 = Bloom, J. S. et al. 2000, GCN 702\\
 Cas511 = Castro-Tirado, A. J. et al. 1999, ApJ, 511, L85\\
 Djo139 = Djorgovski, S. G. et al. 1998, GCN 139\\
 Hol371 = Holland, S. et al. 2001, A\&A, 371, 52\\
 Sok372 = Sokolov, V. V. et al. 2001, A\&A, 372, 438\\
 Vre523 = Vreeswijk, P. M. et al. 1999, ApJ, 523, 171\\
 }
\tablecomments{$\ $\\ $\ $\\
1 = derived from spectrum\\
2 = converted from HST STIS 50CCD Clear measurement\\
}
\end{deluxetable}
\begin{deluxetable}{llllllll}
\tablecaption{\bf{GRB 990123}}
\tablehead{
\colhead{$dt$}  &
\colhead{mag} &
\colhead{error} &
\colhead{band} &
\colhead{date} &
\colhead{corrections} &
\colhead{reference} &
\colhead{remark}}
\startdata
0.3484        & 19.10  & 0.15   & \fltr{R}{C}     & 990123.756     &    NE,NH & Cas283   & \\
0.5194        & 19.42  & 0.17   & \fltr{R}{C}     & 990123.927     &    NE,NH & Cas283   & \\
0.5504        & 19.69  & 0.06   & \fltr{R}{C}     & 990123.958     &    NE,NH & Cas283   & \\
0.5624        & 19.58  & 0.14   & \fltr{R}{C}     & 990123.970     &    NE,NH & Gal398   & \\
0.5924  &   19.61   &   0.03    &   \fltr{R}{C} &   990124.000  &   NE,NH   &   Sag27   &   \\
0.6544        & 19.94  & 0.13   & \fltr{R}{C}     & 990124.062     &    NE,NH & Gal398   & \\
0.6734        & 20.08  & 0.11   & \fltr{R}{C}     & 990124.081     &    NE,NH & Gal398   & \\
0.6884        & 19.89  & 0.04   & \fltr{R}{C}     & 990124.096     &    NE,NH & Cas283   & \\
\enddata
\tablerefs{$\ $\\ $\ $\\
Ake398 = Akerlof, C. et al. 1999, Nature, 398, 400\\
AZ = Zeh, A. et al., unpublished\\
Cas283 = Castro-Tirado, A. J. et al. 1999, Science, 283, 2069\\
Fru519 = Fruchter, A. S. et al. 1999, ApJ, 519, L13\\
Gal398 = Galama, T. J. et al. 1999, Nature, 398, 394\\
Hal257 = Halpern, J. P. et al. 1999, GCN 257\\
Hen307 = Henden, A. et al. 1999, GCN 307\\
Hol344 = Holland, S. et al. 1999, A\&A, 344, L67\\
Kul398 = Kulkarni, S. R. et al. 1999, Nature, 398, 389\\
Lac7096 = Lachaume, R. et al. 1999, IAUC 7096\\
Mau220 = Maury, A. et al. 1999, GCN 220\\
Off7098 = Offutt, W. et al. 1999, IAUC 7098\\
Sag27 = Sagar, R. et al. 1999, BASI, 27, 3\\
Sok372 = Sokolov, V. V. et al. 2001, A\&A, 372, 438\\
Vei253 = Veillet, C. et al. 1999, GCN 253\\
Vei260 = Veillet, C. et al. 1999, GCN 260\\
Yad242 = Yadigaroglu, I. A. et al. 1999, GCN 242\\
 }
\tablecomments{$\ $\\ $\ $\\
1 = Transformed from HST STIS 50CCD Clear image\\
2 = Dimmed by 0.6 mag according to comment of B. A. Skiff, IAUC 7098\\
}
\end{deluxetable}
\begin{deluxetable}{llllllll}
\tablecaption{\bf{GRB 990308}}
\tablehead{
\colhead{$dt$}  &
\colhead{mag} &
\colhead{error} &
\colhead{band} &
\colhead{date} &
\colhead{corrections} &
\colhead{reference} &
\colhead{remark}}
\startdata
0.13931 &   18.14       &   0.06    &   R   &   990308.35814    &   NE,NH   &   Sch524  &   \\
0.14465 &   18.22       &   0.05    &   R   &   990308.36348    &   NE,NH   &   Sch524  &   \\
81.8        &   $>$22.9 &               &   R   &   990529              &   NE,NH   &   Sch524  &   \\
95.8        &   $>$25.4 &               &   R   &   990612              &   NE,NH   &   Sch524  &   \\
102.8       &   $>$25.7 &               &   R   &   990619              &   NE,NH   &   Sch524  &   \\
468         &   29.4        &   0.4     &   R   &   000619              &   NE,NH   &   Jau402  &   1\\
\enddata
\tablerefs{$\ $\\ $\ $\\
 Jau402 = Jaunsen, A. O. et al. 2003, A\&A, 402, 125\\
 Sch524 = Schaefer, B. E. et al. 1999, ApJ, 524, L103\\
 }
\tablecomments{$\ $\\ $\ $\\
1 = most probable host galaxy, transformed from HST STIS 50CCD Clear filter\\
}
\end{deluxetable}
\begin{deluxetable}{llllllll}
\tablecaption{\bf{GRB 990510}}
\tablehead{
\colhead{$dt$}  &
\colhead{mag} &
\colhead{error} &
\colhead{band} &
\colhead{date} &
\colhead{corrections} &
\colhead{reference} &
\colhead{remark}}
\startdata
0.1466         & 17.84   & 0.02    & V       & 990510.514      & NE,NH & Har523 & \\
0.1546         & 17.88   & 0.02    & V       & 990510.522      & NE,NH & Har523 & \\
0.1616         & 17.95   & 0.01    & V       & 990510.529      & NE,NH & Har523 & \\
0.4076         & 18.84   & 0.06    & V       & 990510.775      & NE,NH & Har523 & \\
0.4156         & 18.90   & 0.08    & V       & 990510.783      & NE,NH & Har523 & \\
0.4236         & 18.98   & 0.05    & V       & 990510.791      & NE,NH & Har523 & \\
0.6116         & 19.23   & 0.04    & V       & 990510.979      & NE,NH & Har523 & \\
0.6436         & 19.39   & 0.05    & V       & 990511.011      & NE,NH & Har523 & \\
\enddata
\tablerefs{$\ $\\ $\ $\\
 Beu352 = Beuermann, K. et al. 1999, A\&A, 352, L26\\
 Blo756 = Bloom, J. S. et al. 2000, GCN 756\\
 Fru386 = Fruchter, A. et al. 1999, GCN 386\\
 Fru757 = Fruchter, A. et al. 2000, GCN 757\\
 Har523 = Harrison, F. A. et al. 1999, ApJ, 523, L121\\
 Pie316 = Pietrzynski, G. et al. 1999, GCN 316\\
 (http://www.astrouw.edu.pl/\symbol{126}ogle/ogle2/GRB990510.html)\\
 Sta522 = Stanek, K. Z. et al. 1999, ApJ, 522, L39\\
 }
\tablecomments{$\ $\\ $\ $\\
Hjorth, J. 1999 (GCN \# 320) report 31 \fltr{R}{C} images with the Danish 1.54m telescope
and present their light curve on a webpage but do not give the data points.\\
1 = converted from HST STIS 50CCD Clear measurements\\
}
\end{deluxetable}
\begin{deluxetable}{llllllll}
\tablecaption{\bf{GRB 990705}}
\tablehead{
\colhead{$dt$}  &
\colhead{mag} &
\colhead{error} &
\colhead{band} &
\colhead{date} &
\colhead{corrections} &
\colhead{reference} &
\colhead{remark}}
\startdata
0.2773        & 16.57  & 0.05   & H      & 990705.945     & NE,NH & Mas354 & \\
0.7483        & 18.38  & 0.05   & H      & 990706.416     & NE,NH & Mas354 & \\
1.2873        & $>$19.9& 0      & H      & 990706.955     & NE,NH & Mas354 & \\
\enddata
\tablerefs{$\ $\\ $\ $\\
 Mas354 = Masetti, N. et al. 2000, A\&A, 354, 473\\
}
\end{deluxetable}
\begin{deluxetable}{llllllll}
\tablecaption{\bf{GRB 990712}}
\tablehead{
\colhead{$dt$}  &
\colhead{mag} &
\colhead{error} &
\colhead{band} &
\colhead{date} &
\colhead{corrections} &
\colhead{reference} &
\colhead{remark}}
\startdata
0.176         & 19.349 & 0.019  & \fltr{R}{C}     & 990712.873     & NE,NH & Sah540  & \\
0.432         & 20.175 & 0.011  & \fltr{R}{C}     & 990713.129     & NE,NH & Sah540  & \\
0.454         & 20.183 & 0.009  & \fltr{R}{C}     & 990713.151     & NE,NH & Sah540  & \\
0.477         & 20.251 & 0.085  & \fltr{R}{C}     & 990713.174     & NE,NH & Sah540  & \\
0.631         & 20.470 & 0.10   & \fltr{R}{C}     & 990713.328     & NE,NH & Sah540  & \\
0.686         & 20.533 & 0.016  & \fltr{R}{C}     & 990713.383     & NE,NH & Sah540  & \\
0.698         & 20.460 & 0.057  & \fltr{R}{C}     & 990713.395     & NE,NH & Sah540  & \\
1.053         & 20.857 & 0.088  & \fltr{R}{C}     & 990713.750     & NE,NH & Sah540  & \\
\enddata
\tablerefs{$\ $\\ $\ $\\
 Fru752 = Fruchter, A. et al. 2000, GCN 752\\
 Hjo534 = Hjorth, J. et al. 2000, ApJ, 534, L147\\
 Sah540 = Sahu, K. C. et al. 2000, ApJ, 540, 74\\
 }
\tablecomments{$\ $\\ $\ $\\
1 = date given only as "September 2000" - middle of the month assumed\\
}
\end{deluxetable}
\begin{deluxetable}{llllllll}
\tablecaption{\bf{GRB 991208}}
\tablehead{
\colhead{$dt$}  &
\colhead{mag} &
\colhead{error} &
\colhead{band} &
\colhead{date} &
\colhead{corrections} &
\colhead{reference} &
\colhead{remark}}
\startdata
0.0001157   &   $>$4    &   0   &   \fltr{R}{C} &   991208.1917824  &   NE,NH   &   Cas370  &   \\
2.0785        & 18.7   & 0.1    & \fltr{R}{C}     & 991210.2708    & NE,NH  &   Cas370  &  \\
3.0188        & 19.60  & 0.03   & \fltr{R}{C}     & 991211.2111    & NE,NH  &   Cas370  &  \\
3.0584        & 19.61  & 0.04   & \fltr{R}{C}     & 991211.2507    & NE,NH  &   Cas370  &  \\
3.0869        & 19.70  & 0.08   & \fltr{R}{C}     & 991211.2792    & NE,NH  &   Cas370  &  \\
4.0258        & 20.0   & 0.3    & \fltr{R}{C}     & 991212.2181    & NE,NH  &   Cas370  &  \\
4.0577        & 19.9   & 0.5    & \fltr{R}{C}     & 991212.2500    & NE,NH  &   Cas370  &  \\
4.0681        & 20.37  & 0.05   & \fltr{R}{C}     & 991212.2604    & NE,NH  &   Cas370  &  \\
\enddata
\tablerefs{$\ $\\ $\ $\\
 AZ = Zeh, A. et al., unpublished\\
 Cas370 = Castro-Tirado, A. J. et al. 2001, A\&A, 370, 398\\
 Dod461 = Dodonov, S. et al. 1999, GCN 461\\
 Gar456 = Garnavich, P. et al. 1999, GCN 456\\
 Hal458 = Halpern, J. P. et al. 1999, GCN 458\\
}
\end{deluxetable}
\begin{deluxetable}{llllllll}
\tablecaption{\bf{GRB 991216}}
\tablehead{
\colhead{$dt$}  &
\colhead{mag} &
\colhead{error} &
\colhead{band} &
\colhead{date} &
\colhead{corrections} &
\colhead{reference} &
\colhead{remark}}
\startdata
0.4485        & 18.49  & 0.05   & \fltr{R}{C}     & 991217.12      & NE,NH & Hal543  &  \\
0.4675        & 18.59  & 0.06   & \fltr{R}{C}     & 991217.139     & NE,NH & Hal543  &  \\
0.4765        & 18.63  & 0.02   & \fltr{R}{C}     & 991217.148     & NE,NH & Dol486  &  \\
0.4775        & 18.60  & 0.06   & \fltr{R}{C}     & 991217.149     & NE,NH & Hal543  &  \\
0.4805        & 18.64  & 0.02   & \fltr{R}{C}     & 991217.152     & NE,NH & Dol486  &  \\
0.4815        & 18.52  & 0.08   & \fltr{R}{C}     & 991217.153     & NE,NH & Hal543  &  \\
0.4815        & 18.7  & 0.2   & \fltr{R}{C}     & 991217.153     & NE,NH & How479    &  \\
0.4855        & 18.64  & 0.06   & \fltr{R}{C}     & 991217.157     & NE,NH & Hal543  & 1 \\
\enddata
\tablerefs{$\ $\\ $\ $\\
 Dol486 = Dolan, C. et al. 1999, GCN 486\\
 Gar543 = Garnavich, P. M. et al. 2000, ApJ, 543, 61\\
 Hal543 = Halpern, J. P. et al. 2000, ApJ, 543, 697\\
 Hen473 = Henden, A. et al. 1999, GCN 473\\
 How479 = Howell, D. A. et al. 1999, GCN 479\\
 Sag28 = Sagar, R. et al. 2000, BASI, 28, 15\\
 Sch517 = Schaefer, B. E. et al. 2000, GCN 517\\
 Vre751 = Vreeswijk, P. M. et al. 2000, GCN 751\\
 }
\tablecomments{$\ $\\ $\ $\\
1 = different measurements\\
2 = converted from HST STIS CL and LP measurement, $0.\!\arcsec4$ aperature\\
}
\end{deluxetable}
\begin{deluxetable}{llllllll}
\tablecaption{\bf{GRB 000131}}
\tablehead{
\colhead{$dt$}  &
\colhead{mag} &
\colhead{error} &
\colhead{band} &
\colhead{date} &
\colhead{corrections} &
\colhead{reference} &
\colhead{remark}}
\startdata
3.5134  &   23.26           &   0.04    &   R                       &   000204.137  &   NE,NH   &   And364  &   \\
5.5594  &   24.35           &   0.14    &   R                       &   000206.183  &   NE,NH   &   And364  &   \\
7.4674  &   25.13           &   0.19    &   R                       &   000208.091  &   NE,NH   &   And364  &   \\
33.4014 &   $>$25.70    &               &   R                       &   000305.025  &   NE,NH   &   And364  &   \\
\enddata
\tablerefs{$\ $\\ $\ $\\
 And364 = Andersen, M. I. et al. 2000, A\&A, 364, L54\\
 }
\tablecomments{$\ $\\ $\ $\\
GCN 1133 (Bloom, J. S. et al. 2001) concerns HST detection of the host
565 days after the burst in F606W and F814W filters. No magnitude is given.
Results presented at a conference (Fox, D. B. et al. 2002, Bibliographic Code:
2002APS..APRN17050F) are not accessible.}
\end{deluxetable}
\begin{deluxetable}{llllllll}
\tablecaption{\bf{GRB 000301C}}
\tablehead{
\colhead{$dt$}  &
\colhead{mag} &
\colhead{error} &
\colhead{band} &
\colhead{date} &
\colhead{corrections} &
\colhead{reference} &
\colhead{remark}}
\startdata
1.5189        & 20.42  & 0.06   & \fltr{R}{C}     & 000302.93      & NE,NH & Sag29  &  \\
1.5192        & 20.42  & 0.04   & \fltr{R}{C}     & 000302.9303    & NE,NH & Mas359 &  \\
1.550         & 20.02  & 0.03   & \fltr{R}{C}     & 000302.9609    & NE,NH & Rho546 &  \\
1.5509        & 20.02  & 0.028  & \fltr{R}{C}     & 000302.9618    & NE,NH & Bha545 &  \\
1.7291        & 20.09  & 0.04   & \fltr{R}{C}     & 000303.14      & NE,NH & Jen370 &  \\
1.7347        & 20.25  & 0.05   & \fltr{R}{C}     & 000303.1457    & NE,NH & Mas359 &  \\
1.7591        & 20.15  & 0.04   & \fltr{R}{C}     & 000303.17      & NE,NH & Jen370 &  \\
1.7791        & 20.11  & 0.04   & \fltr{R}{C}     & 000303.19      & NE,NH & Jen370 &  \\
\enddata
\tablerefs{$\ $\\ $\ $\\
 Bha545 = Bhargavi, S. G. 2000, ApJ, 545, L77\\
 Fru1063 = Fruchter, A. S. 2001, GCN 1063\\
 Jen370 = Jensen, B. L. 2001, A\&A, 370, 909\\
 Mas359 = Masetti, N. 2000, A\&A, 359, L23\\
 Rho546 = Rhoads, J. E. 2001, ApJ, 546, 117\\
 Sag28 = Sagar, R. 2000, BASI, 28, 499\\
 Sok372 = Sokolov, V. V. 2001, A\&A, 372, 438\\
 Vei623 = Veillet, C. 2000,  GCN 623\\
 }
\tablecomments{$\ $\\ $\ $\\
1 = Date given as 010211 on http://www.ifa.au.dk/\symbol{126}hst/grb\_hosts/data/grb000301C/index.html\\
}
\end{deluxetable}
\begin{deluxetable}{llllllll}
\tablecaption{\bf{GRB 000418}}
\tablehead{
\colhead{$dt$}  &
\colhead{mag} &
\colhead{error} &
\colhead{band} &
\colhead{date} &
\colhead{corrections} &
\colhead{reference} &
\colhead{remark}}
\startdata
2.48      & 21.63   & 0.04 & \fltr{R}{C}     & 000420.89      & NE,NH & Klo545   & \\
2.53      & $>$20.5   & 0 & \fltr{R}{C}     & 000420.94      & NE,NH & Klo545   & \\
2.738     & 21.66   & 0.12 & \fltr{R}{C}     & 000421.15       & CE,NH & Ber556   & \\
2.74      & 21.77   & 0.12 & \fltr{R}{C}     & 000421.15      & NE,NH & Klo545   & \\
3.45      & 22.01   & 0.14 & \fltr{R}{C}     & 000421.86      & NE,NH & Klo545   &     \\
7.904     & 22.65   & 0.20 & \fltr{R}{C}     & 000426.316     & CE,NH & Ber556   & \\
7.91      & 22.74   & 0.20 & \fltr{R}{C}     & 000426.32      & NE,NH & Klo545   & \\
8.848     & 22.77   & 0.23 & \fltr{R}{C}     & 000427.26       & CE,NH & Ber556   & \\
\enddata
\tablerefs{$\ $\\ $\ $\\
 AZ = Zeh, A. et al., unpublished\\
 Ber556 = Berger, E. et al. 2001, ApJ, 556, 556\\
 Fru1061 = Fruchter, A. et al. 2001, GCN 1061\\
 Gor409 = Gorosabel, J. et al. 2003, A\&A, 409, 123\\
 Klo545 = Klose, S. et al. 2000, ApJ, 545, 271\\
 Met733 = Metzger, M. et al. 2000, GCN 733;\\
 }
\tablecomments{$\ $\\ $\ $\\
1 = converted from HST STIS 50CCD Clear measurement\\
}
\end{deluxetable}
\begin{deluxetable}{llllllll}
\tablecaption{\bf{GRB 000630}}
\tablehead{
\colhead{$dt$}  &
\colhead{mag} &
\colhead{error} &
\colhead{band} &
\colhead{date} &
\colhead{corrections} &
\colhead{reference} &
\colhead{remark}}
\startdata
0.85        & 23.00  & 0.30   & \fltr{R}{C}     & 000630.87      & NE,NH & Fyn369  &  \\
0.88        & 23.04  & 0.08   & \fltr{R}{C}     & 000630.90      & NE,NH & Fyn369  &  \\
1.22        & 23.13  & 0.25   & \fltr{R}{C}     & 000701.24      & NE,NH & Fyn369  &  \\
1.2726        & 23.47  & 0.13   & \fltr{R}{C}     & 000701.2926      & NE,NH & Yos748  &  \\
1.3152        & $>$23.2  & 0   & \fltr{R}{C}     & 000701.3352      & NE,NH & Hal745  &  \\
1.86        & 24.05  & 0.16   & \fltr{R}{C}     & 000701.88      & NE,NH & Fyn369  &  \\
1.87        & 23.85  & 0.22   & \fltr{R}{C}     & 000701.89      & NE,NH & Fyn369  &  \\
3.89        & 24.67  & 0.14   & \fltr{R}{C}     & 000703.91      & NE,NH & Fyn369  &  \\
10.88        & $>$25.3& 0      & \fltr{R}{C}     & 000710.9       & NE,NH & Fyn369  &  \\
25.88        & $>$26.1& 0      & \fltr{R}{C}     & 000725.9       & NE,NH & Fyn369  &  \\
355            & 26.68  & 0.2    & \fltr{R}{C}     & 010619         & NE,NH & Kap1069 &    \\
\enddata
\tablerefs{$\ $\\ $\ $\\
 Fyn369 = Fynbo, J. U. et al. 2001, A\&A, 369, 373\\
 Hal745 = Halpern, J. P. et al. 2000, GCN 745\\
 Kap1069 = Kaplan, D. L. et al. 2001, GCN 1069\\
 Yos748 = Yost, S. et al. 2000, GCN 748\\
 }
\end{deluxetable}
\begin{deluxetable}{llllllll}
\tablecaption{\bf{GRB 000911}}
\tablehead{
\colhead{$dt$}  &
\colhead{mag} &
\colhead{error} &
\colhead{band} &
\colhead{date} &
\colhead{corrections} &
\colhead{reference} &
\colhead{remark}}
\startdata
1.435   & 20.700& 0.081& \fltr{R}{C} & 000912.737 & Pri573 &  \\
1.435   & 20.70 & 0.08 & \fltr{R}{C} & 000912.737 & Laz378 &  \\
1.435   & 20.57 & 0.08 & \fltr{R}{C} & 000912.737 & Mas438 & 1 \\
3.795   & $>$20.5 & 0  & \fltr{R}{C} & 000915.097 & Ped815 &  \\
4.291   & 22.299& 0.030& \fltr{R}{C} & 000915.593 & Pri573 &  \\
4.30    & 22.30 & 0.06 & \fltr{R}{C} & 000915.602 & Laz378 &  \\
4.30    & 22.17 & 0.06 & \fltr{R}{C} & 000915.602 & Mas438 & 1 \\
4.933   & 22.42 & 0.15 & \fltr{R}{C} & 000916.235 & Mas438 &  \\
\enddata
\tablerefs{$\ $\\ $\ $\\
 Laz378 = Lazzati, D. et al. 2001, A\&A, 378, 996\\
 Mas438 = Masetti, N. et al. 2005, A\&A, 438, 841\\
 Ped815 = Pedersen, H. et al. 2000, GCN 815\\
 Pri573 = Price, P. A. et al. 2002, ApJ, 573, 85\\
 }
\tablecomments{$\ $\\ $\ $\\
1 = Laz378 data transformed to Mas438 zero point:\\
$\Delta$mag(B)=0.19mag, $\Delta$mag(V)=0.22mag, $\Delta$mag(\fltr{R}{C})=0.13mag, $\Delta$mag(\fltr{I}{C})=0.22mag\\
}
\end{deluxetable}
\begin{deluxetable}{lllllllll}
\tablecaption{\bf{GRB 000926}}
\tablehead{
\colhead{$dt$}  &
\colhead{mag} &
\colhead{+error} &
\colhead{-error} &
\colhead{band} &
\colhead{date} &
\colhead{corrections} &
\colhead{reference} &
\colhead{remark}}
\startdata
2.3778  &   21.23   &   0.03    &   0.03    &   R   &   000929.3075 &   NE,NH   &   Vei831  &   \\
3.3778  &   22.10   &   0.04    &   0.04    &   R   &   000930.3075 &   NE,NH   &   Vei831  &   \\
4.3296  &   22.71   &   0.04    &   0.04    &   R   &   001001.2593 &   NE,NH   &   Vei831  &   \\
5.3329  &   23.13   &   0.08    &   0.08    &   R   &   001002.2626 &   NE,NH   &   Vei831  &   \\
6.3007  &   23.45   &   0.10    &   0.10    &   R   &   001003.2304 &   NE,NH   &   Vei831  &   \\
7.3203  &   23.73   &   0.15    &   0.15    &   R   &   001004.2500 &   NE,NH   &   Vei831  &   \\
&&&&&&&&\\
0.925     & 19.326  &   0.015   &   0.015   &   \fltr{R}{C} &   000927.8547 &   NE,CH   &   Fyn373  &   \\
\enddata
\tablerefs{$\ $\\ $\ $\\
 Cas586 = Castro, S. et al. 2003, ApJ, 586, 129\\
 Fat30 = Fatkhullin, T. A. et al. 2004, Astronomy Letters, 2004, Vol. 30, 283\\
 Fyn373 = Fynbo, J. U. et al. 2001, A\&A, 373, 796\\
 Har559 = Harrison, F. A. et al. 2001, ApJ, 559, 123\\
 Pri549 = Price, P. A. et al. 2001, ApJ, 549, L7\\
 Sag29 = Sagar, R. et al. 2001, BASI, 29, 1\\
 Vei831 = Veillet, C. et al. 2000, GCN 831\\
 Vrb819 = Vrba, J. et al. 2000, GCN 819\\
 }
\tablecomments{$\ $\\ $\ $\\
1 = converted from HST WFPC2 measurements, flux density in $\mu$Jy\\
2 = converted from HST WFPC2 measurements, 2 pixel ($0.\!\arcsec25$) radius around OT position\\
3 = converted from HST WFPC2 measurements, 5 pixel ($0.\!\arcsec625$) radius around OT position\\
4 = re-reddened with $E_{(B-V)}=0.024$\\
}
\end{deluxetable}
\begin{deluxetable}{llllllll}
\tablecaption{\bf{GRB 001007}}
\tablehead{
\colhead{$dt$}  &
\colhead{mag} &
\colhead{error} &
\colhead{band} &
\colhead{date} &
\colhead{corrections} &
\colhead{reference} &
\colhead{remark}}
\startdata
3.4825   &  20.3        &           &   R   &   001010.6900 &   NE,NH   &   Pri843  &   \\
3.9502   &  20.5        &   0.13    &   R   &   001011.1577 &   NE,NH   &   Cas393  &   \\
4.1358   &  20.77       &   0.06    &   R   &   001011.3433 &   NE,NH   &   Cas393  &   \\
5.0504   &  $>$20.30    &           &   R   &   001012.2579 &   NE,NH   &   Cas393  &   \\
13.0327  &  $>$22.10    &           &   R   &   001020.2402 &   NE,NH   &   Cas393  &   \\
18.1525  &  $>$21.6   &         &   R   &   001025.3600 &   NE,NH   &   Pri862  &   \\
22.0754  &  23.98       &   0.17    &   R   &   001029.2829 &   NE,NH   &   Cas393  &   \\
25.9447  &  24        & 0.16    &   R   &   001102.1522 &   NE,NH   &   Cas393  &   \\
39.9386  &  24.53       &   0.22    &   R   &   001116.1461 &   NE,NH   &   Cas393  &   \\
467.8631 &  24.84       &   0.15    &   R   &   020118.0706 &   NE,NH   &   Cas393  &   \\
\enddata
\tablerefs{$\ $\\ $\ $\\
 Cas393 = Castro Cer$\acute{\tn{o}}$n, J. M. et al. 2002, A\&A, 393, 445\\
 Pri843 = Price, P. A. et al. 2000, GCN 843\\
 Pri862 = Price, P. A. et al. 2000, GCN 862\\
}
\end{deluxetable}
\begin{deluxetable}{llllllll}
\tablecaption{\bf{GRB 001011}}
\tablehead{
\colhead{$dt$}  &
\colhead{mag} &
\colhead{error} &
\colhead{band} &
\colhead{date} &
\colhead{corrections} &
\colhead{reference} &
\colhead{remark}}
\startdata
0.3516      &   20.99       &   0.05    &   R                       &   001012.0148 &   NE,NH   &   Gor384  &   \\
2.3644      &   23.74       &   0.23    &   R                       &   001014.0275 &   NE,NH   &   Gor384  &   \\
158.7256    &   $>$24.3 &               &   R                       &   010419.3887 &   NE,NH   &   Gor384  &   \\
159.7313    &   $>$24.6 &               &   R                       &   010420.3945 &   NE,NH   &   Gor384  &   \\
161.7314    &   $>$24.5 &               &   R                       &   010422.3945 &   NE,NH   &   Gor384  &   \\
189.513     &   25.38       &   0.25    &   R                       &   010520.1762 &   NE,NH   &   Gor384  &   \\
\enddata
\tablerefs{$\ $\\ $\ $\\
 Gor384 = Gorosabel, J. et al. 2002, A\&A, 384, 11\\
 }
\end{deluxetable}
\begin{deluxetable}{lllllllll}
\tablecaption{\bf{GRB 010222}}
\tablehead{
\colhead{$dt$}  &
\colhead{mag} &
\colhead{+error} &
\colhead{-error} &
\colhead{band} &
\colhead{date} &
\colhead{corrections} &
\colhead{reference} &
\colhead{remark}}
\startdata
0.1515  &   18.40   &   0.05    &   0.05    &   \fltr{R}{C} &   010222.4595 &   NE,NH   &   Sta563  &   \\
0.1530  &   18.36   &   0.03    &   0.03    &   \fltr{R}{C} &   010222.4610 &   NE,NH   &   Sta563  &   \\
0.1615  &   18.40   &   0.03    &   0.03    &   \fltr{R}{C} &   010222.4695 &   NE,NH   &   Sta563  &   \\
0.174   &   18.455  &   0.033   &   0.033   &   \fltr{R}{C} &   010222.482  &   NE,NH   &   Gal587  &   \\
0.179   &   18.481  &   0.036   &   0.036   &   \fltr{R}{C} &   010222.487  &   NE,NH   &   Gal587  &   \\
0.1960  &   18.589  &   0.040   &   0.040   &   \fltr{R}{C} &   010222.5040 &   NE,NH   &   Wat53   &   \\
0.2000  &   18.4    &   0.1 &   0.1 &   \fltr{R}{C} &   010222.508  &   NE,NH   &   Cow29   &   \\
0.2005  &   18.590  &   0.030   &   0.030   &   \fltr{R}{C} &   010222.5085 &   NE,NH   &   Wat53   &   \\
\enddata
\tablerefs{$\ $\\ $\ $\\
 Cow29 = Cowsik, R. et al. 2001, BASI, 29, 157\\
 Gal587 = Galama, T. J. et al. 2003, ApJ 587, 135\\
 Mas374 = Masetti, N. et al. 2001, A\&A, 374, 382\\
 Oks30 = Oksanen, A. et al. 2002, JAAVSO Volume 30, 126\\
 Oro976 = Orosz, J. A. et al. 2001, GCN 976\\
 Sag29 = Sagar, R. et al. 2001, BASI, 29, 91\\
 Sal1082 = Salamanca, I. et al. 2001, GCN 1082\\
 Sta563 = Stanek, K. Z. et al. 2001, ApJ, 563, 592\\
 Uem984 = Uemura, M. et al. 2001, GCN 984\\
 Val986 = Valentini, G. et al. 2001, GCN 986\\
 Wat53 = Watanabe, J. et al. 2001, PASJ, 53, L27\\
 }
\tablecomments{$\ $\\ $\ $\\
1 = HST Filter Magnitudes transformed to Johnson-Cousins Filter System\\
3 = host galaxy corrected for optical transient\\
}
\end{deluxetable}
\begin{deluxetable}{llllllll}
\tablecaption{\bf{GRB 010921}}
\tablehead{
\colhead{$dt$}  &
\colhead{mag} &
\colhead{error} &
\colhead{band} &
\colhead{date} &
\colhead{corrections} &
\colhead{reference} &
\colhead{remark}}
\startdata
1.074   &   19.669  &   0.06    &   \fltr{R}{C} &   010922.293  &   NE,NH   &   Pri571  &   \\
1.074   &   39.116  &   5.072   &   \fltr{R}{C} &   010922.293  &   NE,CH   &   Pri584  &   1\\
1.102   &   19.79   &   0.069   &   \fltr{R}{C} &   010922.321  &   NE,NH   &   Pri571  &   \\
1.102   &   36.135  &   4.486   &   \fltr{R}{C} &   010922.321  &   NE,CH   &   Pri584  &   1\\
21.007  &   21.693  &   0.053   &   \fltr{R}{C} &   011012.226  &   NE,NH   &   Pri571  &   \\
28.053  &   21.807  &   0.051   &   \fltr{R}{C} &   011019.272  &   NE,NH   &   Pri571  &   \\
28.053  &   0.916   &   4.284   &   \fltr{R}{C} &   011019.272  &   NE,CH   &   Pri584  &   1\\
56.7    &   21.93   &   0.09    &   \fltr{R}{C} &   011117.7    &   NE,NH   &   Par571  &   \\
\enddata
\tablerefs{$\ $\\ $\ $\\
 AZ = Zeh, A. et al., unpublished\\
 Par571 = Park, H. S. et al. 2002, ApJ, 571, L131\\
 Pri571 = Price, P. A. et al. 2002, ApJ, 571, L121\\
 Pri584 = Price, P. A. et al. 2003, ApJ, 584, 936\\
 }
\tablecomments{$\ $\\ $\ $\\
1 = flux in $\mu$Jy\\
}
\end{deluxetable}
\begin{deluxetable}{llllllll}
\tablecaption{\bf{GRB 011121}}
\tablehead{
\colhead{$dt$}  &
\colhead{mag} &
\colhead{error} &
\colhead{band} &
\colhead{date} &
\colhead{corrections} &
\colhead{reference} &
\colhead{remark}}
\startdata
0.4323  &   19.06           &   0.03    &   R   &   011122.2152 &   NE,NH   &   Gar582  &   \\
0.4388  &   19.09           &   0.03    &   R   &   011122.2217 &   NE,NH   &   Gar582  &   \\
0.4888  &   19.26           &   0.04    &   R   &   011122.2717 &   NE,NH   &   Gar582  &   \\
0.4970  &   19.31           &   0.03    &   R   &   011122.2799 &   NE,NH   &   Gar582  &   \\
0.5231  &   19.45           &   0.03    &   R   &   011122.3060 &   NE,NH   &   Gar582  &   \\
0.5439  &   19.50           &   0.03    &   R   &   011122.3268 &   NE,NH   &   Gar582  &   \\
0.5716  &   19.59           &   0.03    &   R   &   011122.3545 &   NE,NH   &   Gar582  &   \\
1.5323  &   21.41           &   0.05    &   R   &   011123.3152 &   NE,NH   &   Gar582  &   \\
\enddata
\tablerefs{$\ $\\ $\ $\\
 Blo572 = Bloom, J. S. et al. 2002, ApJ, 572, L45\\
 Gar582 = Garnavich, P. M. et al. 2003, ApJ, 582, 924\\
 Gre599 = Greiner, J. et al. 2003, ApJ, 599, 1223\\
}
\end{deluxetable}
\begin{deluxetable}{llllllll}
\tablecaption{\bf{GRB 011211}}
\tablehead{
\colhead{$dt$}  &
\colhead{mag} &
\colhead{error} &
\colhead{band} &
\colhead{date} &
\colhead{corrections} &
\colhead{reference} &
\colhead{remark}}
\startdata
0.4199  &   20.012  &   0.033   &   R   &   011212.2181 &   NE,NH   &   JakNA9  &   \\
0.4245  &   19.877  &   0.029   &   R   &   011212.2227 &   NE,NH   &   JakNA9  &   \\
0.429       &   19.918  &   0.022   &   R   &   011212.2272 &   NE,NH   &   JakNA9  &   \\
0.45        &   20.09       &   0.07    &   R   &   011212.2482 &   NE,NH   &   Hol124  &   \\
0.4518  &   20.28       &   0.02    &   R   &   011212.2500 &   NE,NH   &   Hol124  &   \\
0.4585  &   20.4        &   0.04    &   R   &   011212.2567 &   NE,NH   &   Hol124  &   \\
0.4789  &   20.37       &   0.04    &   R   &   011212.2771 &   NE,NH   &   Hol124  &   \\
0.4849  &   20.368  &   0.04    &   R   &   011212.2831 &   NE,NH   &   JakNA9  &   \\
\enddata
\tablerefs{$\ $\\ $\ $\\
 Hol124 = Holland, S. T. et al. 2002, AJ, 124, 639\\
 Jak408 = Jakobsson, P. et al. 2003, A\&A, 408, 941\\
 JakNA9 = Jakobsson, P. et al. 2004, NA, 9, 435\\
}
\end{deluxetable}
\begin{deluxetable}{lllllllll}
\tablecaption{\bf{GRB 020124}}
\tablehead{
\colhead{$dt$}  &
\colhead{mag} &
\colhead{+ error} &
\colhead{- error} &
\colhead{band} &
\colhead{date} &
\colhead{corrections} &
\colhead{reference} &
\colhead{remark}}
\startdata
0.06985 &   18.219  &   0.046   &   0.046   &   R                       &   020124.51516    &   NE,NH   &   Ber581  &   \\
0.072       &   $>$17.8 &               &               &   R                       &   020100.130      &   NE,NH   &   Kaw1343 &   \\
0.07842 &   18.376  &   0.082   &   0.082   &   R                       &   020124.52373    &   NE,NH   &   Ber581  &   \\
0.246       &   $>$18.6 &               &               &   R                       &   020100.150      &   NE,NH   &   Kaw1343 &   \\
0.281       &   $>$18.8 &               &               &   R                       &   020100.000      &   NE,NH   &   Kaw1343 &   \\
0.323       &   $>$19.1 &               &               &   R                       &   020100.000      &   NE,NH   &   Kaw1343 &   \\
0.364       &   $>$19.1 &               &               &   R                       &   020100.517      &   NE,NH   &   Kaw1343 &   \\
0.406       &   $>$17.7 &               &               &   R                       &   020100.691      &   NE,NH   &   Kaw1343 &   \\
\enddata
\tablerefs{$\ $\\ $\ $\\
 Ber581 = Berger, E. et al. 2002, ApJ, 581, 981\\
 Hjo597 = Hjorth, J. et al. 2003, ApJ, 597, 699\\
 Kaw1343 = Kawai, N. et al. 2002, GCN 1343\\
  }
\tablecomments{$\ $\\ $\ $\\
1 = MACHO Filter\\
2 = converted from HST STIS 50CCD Clear Filter\\
3 = co-addition of 24.855 and 32.265 days\\
}
\end{deluxetable}
\begin{deluxetable}{lllllllll}
\tablecaption{\bf{GRB 020305}}
\tablehead{
\colhead{$dt$}  &
\colhead{mag} &
\colhead{error+} &
\colhead{error-} &
\colhead{band} &
\colhead{date} &
\colhead{corrections} &
\colhead{reference} &
\colhead{remark}}
\startdata
0.67        &   $>$18       &           &   &   R                   &   020306.17   &   NE,NH   &   Mor1264 &   \\
3.84        &   $>$21.5 &           &   &   R                   &   020309.34   &   NE,NH   &   Pri1267 &   \\
7.1         &   22.6        &           &   &   R                   &   020312.6    &   NE,NH   &   Ohy1271 &   \\
8.3         &   22.1        &           &   &   R                   &   020313.8    &   NE,NH   &   Fox1270 &   \\
11.5192 &   22.76   &   0.05    &   0.05    &   R   &   020317.0160 &   NE,NH   &   Gor437  &   \\
13.5032 &   22.48   &   0.08    &   0.08    &   R   &   020319.0000 &   NE,NH   &   Gor437  &   \\
15.5817 &   22.42   &   0.06    &   0.06    &   R   &   020321.0785 &   NE,NH   &   Gor437  &   \\
16.5837 &   22.41   &   0.05    &   0.05    &   R   &   020322.0805 &   NE,NH   &   Gor437  &   \\
\enddata
\tablerefs{$\ $\\ $\ $\\
 Fox1270 = Fox, D. W. et al. 2002, GCN 1270\\
 Gor437 = Gorosabel, J. et al. 2005, A\&A, 437, 411\\
 Lee1275 = Lee, B. C. et al. 2002, GCN 1275\\
 Mor1264 = Moran, J. et al. 2002, GCN 1264\\
 Ohy1271 = Ohyama, Y. et al. 2002, GCN 1271\\
 Pri1261 = Price, P. A. et al. 2002, GCN 1261\\
 Pri1267 = Price, P. A. et al. 2002, GCN 1267\\
  }
\tablecomments{$\ $\\ $\ $\\
Note: The location of the optical afterglow discovered in GCN 1267
lies outside the image reported in GCN 1265 (Jelinek, M. et al. 2002),
thus their upper limit is not included.\\ R magnitude from Fox1270
(Palomar $200''$) is too bright. According to GCN, Gunn i Filter was used,
but magnitude is given as R. GCN 1542 (Gorosabel, J. et al. -- link to webpage)
also lists data point as R.
}
\end{deluxetable}
\begin{deluxetable}{llllllll}
\tablecaption{\bf{GRB 020322}}
\tablehead{
\colhead{$dt$}  &
\colhead{mag} &
\colhead{error} &
\colhead{band} &
\colhead{date} &
\colhead{corrections} &
\colhead{reference} &
\colhead{remark}}
\startdata
0.279   &   $>$22.5 &               &   R                       &   020322.440  &   NE,NH   &   Sze1295 &   \\
0.283   &   $>$20.8 &               &   R                       &   020322.444  &   NE,NH   &   Fox1292 &   \\
0.283   &   23.26       &   0.32    &   R                       &   020322.444  &   NE,NH   &   Blo1294 &   \\
0.38    &   $>$22.1 &               &   R                       &   020322.54       &   NE,NH   &   Fox1292 &   \\
0.839   &   23.8        &   0.3     &   R                       &   020323.000  &   NE,NH   &   Gre1298 &   \\
1.26    &   $>$24.5 &   0.3     &   R                       &   020323.42       &   NE,NH   &   Wil1307 &   \\
16.8    &   27          &               &   R                       &   020408          &   NE,NH   &   Bur1536 &   \\
16.8    &   $>$30.1 &               &   R                       &   020408          &   NE,CH   &   Bur1536 &   \\
74.8    &   27          &               &   R                       &   020605          &   NE,NH   &   Bur1536 &   \\
            &                   &               &                           &                           &               &                   &   \\
1.248   &   24.5        &   0.7     &   $r^*$                   &   020323.409  &   NE,NH   &   McM1299 &   \\
\enddata
\tablerefs{$\ $\\ $\ $\\
 Blo1294 = Bloom, J. S. et al. 2002, GCN 1294\\
 Bur1536 = Burud, I. et al. 2002, GCN 1536\\
 Fox1292 = Fox, D. W. et al. 2002, GCN 1292\\
 Gre1298 = Greiner, J. et al. 2002, GCN 1298\\
 McM1299 = McMillan, R. J. et al. 2002, GCN 1299\\
 Sze1295 = Szentgyorgyi, A. et al. 2002, GCN 1295\\
 Wil1307 = Williams, G. G. et al. 2002, GCN 1307\\
}
\end{deluxetable}
\include{GRB020331}
\include{GRB020405}
\include{GRB020410}
\include{GRB020813}
\include{XRF020903}
\include{GRB021004}
\include{GRB021211}
\include{GRB030115}
\include{GRB030131}
\include{GRB030226}
\include{GRB030227}
\include{GRB030323}
\include{GRB030324}
\include{GRB030328}
\include{GRB030329}
\include{GRB030418}
\include{GRB030429}
\include{GRB030528}
\include{XRF030723}
\include{GRB030725}
\include{GRB031203}
\include{GRB031220}
\include{GRB040106}
\include{GRB040422}
\include{GRB040827}
\include{XRF040916}
\include{GRB040924}
\include{GRB041006}


\end{document}